%% file: sn-article.tex
\documentclass[pdflatex,sn-mathphys-num]{sn-jnl}

\usepackage{graphicx}%
\usepackage{multirow}%
\usepackage{amsmath,amssymb,amsfonts}%
\usepackage{amsthm}%
\usepackage{mathrsfs}%
\usepackage[title]{appendix}%
\usepackage{xcolor}%
\usepackage{textcomp}%
\usepackage{manyfoot}%
\usepackage{booktabs}%
\usepackage{algorithm}%
\usepackage{algorithmicx}%
\usepackage{algpseudocode}%
\usepackage{listings}%
\usepackage{xspace}
\theoremstyle{thmstyleone}%

\usepackage{textcomp}

\usepackage{array}

\usepackage{subcaption}
\usepackage{tikz}
\usepackage{siunitx}
\usepackage{caption}

\begin{document}

\title[Article Title]{Design and Fabrication of a lightweight three-lens corrector system for the 2.34-m Vainu Bappu Telescope}

\author*[1,2]{\fnm{Nitish Singh} }\email{nitish.singh@iiap.res.in}

\author[1]{\fnm{S. Sriram} }\email{ssr@iiap.res.in}

\author[1]{\fnm{Ramya Sethuram} }\email{ramyas.itcc@iiap.res.in}

\author[1]{\fnm{Bharat Kumar Yerra} }\email{bharat.yerra@iiap.res.in}

\author[1]{\fnm{Rahuldeb Burman} }

\author[1]{\fnm{G. Nataraj} }

\author[1]{\fnm{C Chethan} }

\author[1]{\fnm{P. Madan Mohan Kemkar} }

\author[1]{\fnm{K Sagayanathan} }

\author[1]{\fnm{Saikat Das} }

\author[1]{\fnm{Francis Xavier Rozario J} }

\affil*[1]{\orgname{Indian institute of Astrophysics}, \orgaddress{\street{Koramangala}, \city{Bengaluru}, \postcode{560 034}, \state{Karnataka}, \country{India}}}

\affil[2]{\orgdiv{Department of Applied Optics and Photonics}, \orgname{University of Calcutta}, \orgaddress{\street{Salt Lake}, \city{Kolkata}, \postcode{700 106}, \state{West Bengal}, \country{India}}}

\abstract{The Vainu Bappu Telescope (VBT) is a 2.34-m reflector,  primarily supported on-axis field of view, offering high-resolution and low-to-medium resolution spectroscopic observations in its prime and Cassegrain configurations. This study presents the design and fabrication of a compact, lightweight, three-element wide-field corrector (WFC) utilizing three spherical lenses to cover a polychromatic wavelength range over a 30$'$ FoV at prime focus. The WFC design was optimized using ZEMAX, ensuring precision in aberrations, tolerances, and atmospheric dispersion. The fabricated lenses met stringent tolerances, with a $\pm$1~mm deviation in radius of curvature and $\pm$2~mm deviation in center thickness. A mechanical mount was developed to integrate all the WFC lenses, and wavefront error testing for the WFC system was performed using ZYGO interferometry, yielding a Wavefront Error of 0.05~$\lambda$. Laboratory performance tests were designed and conducted using a dedicated setup with achromatic lenses and 100~$\mu m$ fiber-coupled polychromatic light source showed a deviation of 0.1~pixel on-axis and 0.5~pixel at the extreme off-axis field compared to the ZEMAX design, demonstrating that the optical performance of WFC is with minimal aberrations across the entire FoV. The successful integration of the WFC at the VBT prime focus will increase the FoV, enabling the multi-fiber, multi-spectrograph setup in 30$'$ field that will facilitate both OMR and Echelle spectrograph to be used on the same night along with the addition of new multi-object spectrograph and an integral field unit instrument. This will mark a significant upgrade for the VBT, broadening its research potential, and expanding its observational versatility.
}


\maketitle

\section{Introduction}\label{sec:intro}

Over the past few decades, advancements in astronomical instrumentation have significantly enhanced the capability of night sky observations through wide-field imaging, multi-object spectroscopy, and integral field spectroscopy. To enable wide-field imaging and multi-object spectroscopy, telescopes require a large field of view (FoV) and a good light-collecting area that ensures uniform optical performance with minimal optical aberration across the entire field. The use of a system of lenses called wide-field corrector (WFC) can significantly improve a telescope’s imaging and spectroscopic capabilities. They help to achieve higher image quality, better spectral resolution, and more accurate astrometry and photometry by balancing the chromatic aberration and uniform image quality from center to edge of the field. In 1934, Ross designed a three-element system for the 200~inch (5~m) Mount Palomar telescope \citep{1935ApJ....81..156R}. Later, in 1974, Wynne refined this concept by developing a three-lens corrector positioned near the prime focus of parabolic telescopes \citep{1974MNRAS.167..189W}. This corrector was designed to correct multiple optical aberrations, such as coma and astigmatism, significantly improving image quality over a wider FoV. Since then, the use of WFC has grown substantially in astronomical applications, particularly for wide-field imaging and spectroscopy.  Several telescopes with diameters under 4 meters have successfully implemented WFCs, demonstrating impressive capabilities. The 2.5~m SDSS telescope, for instance, employs a two-element corrector, offering a 3$^{\circ}$ FoV in the Cassegrain mode and covering a wavelength range from 0.4 to 1~$\mu m$ \citep{2006AJ....131.2332G}. The University of Hawaii 2.2~m telescope, Wide-Field Imager (UHWFI), consisting of six lenses, provides a 30$'$ FoV and a wavelength range of 0.4 to 1.05~$\mu m$ \cite{2006PASP..118..780H}. The 4.01~m Mayall Telescope, equipped with a WFC at prime focus, offers a 1.5$^{\circ}$ corrected FoV with an additional atmospheric dispersion corrector (ADC), covering 0.4 to 1~$\mu m$  \cite{2024AJ....168...95M} \cite{2014SPIE.9151E..1MS}. The 3.9~m Anglo-Australian Telescope (AAT) offers a 2.1$^{\circ}$ corrected FoV with a six-lens system, spanning wavelengths from 0.365 to 1.1014~$\mu m$ \cite{1994ApOpt..33.7362J} \cite{2002MNRAS.333..279L}. The 3.58~m Canada-France-Hawaii Telescope (CFHT) also features a 4 element  WFC, providing a 1$^{\circ}$ corrected FoV but without an ADC, covering wavelengths from 0.35 to 2~$\mu m$ , (\textit{CFHT Observatory Manual})\footnote{\url{https://www.cfht.hawaii.edu/Instruments/ObservatoryManual/index.html}}. The 4MOST instrument, for instance, employs a six-element WFC, providing a 2.5$^{\circ}$ FoV in the Cassegrain configuration and covering a wavelength range from 0.39 to 0.95~$\mu m$, with an integrated ADC included in the design  \cite{2016SPIE.9908E..89A}. 

The Vainu Bappu Telescope (VBT) is a 2.34~m reflecting equatorial telescope located at Vainu Bappu Observatory (VBO), Kavalur, in India.  The VBT has an f/3.25 parabolic primary mirror, which gives a prime focus image scale of 27~arcsec/mm, and an f/13 hyperbolic secondary mirror, which provides a Cassegrain focus image scale of 6.7~arcsec/mm. The VBT achieved its first light on November 2, 1985 (\cite{1992BASI...20..319B}). The latitude (12$^\circ34.6'$ N) of this telescope location is in good proximity to Earth’s equator, allowing access to the celestial objects in the northern and southern sky alomost equally, and it is the only 2-m class telescope in India that provides access to study astronomical objects in the southern sky. Additionally,  its longitudinal  (78$^\circ49.6'$E) position between the southern facilities in Australia and South-America positions it to fill a critical observational gap for celestial and transient events. 

VBT is equipped with two spectrographs that provide spectra of point sources with a range of resolution R$\sim2000$ to R$\sim100000$. A fiber-fed High-resolution Echelle spectrograph (HRES) is coupled at prime focus (\cite{2005JApA...26..331R}), and a slit-fed low-resolution OMR spectrograph (OMRS) is mounted at Cassegrain focus (\cite{1998BASI...26..383P}). Large sky surveys like Gaia (\cite{2016A&A...595A...1G}), LSST (\cite{2019ApJ...873..111I}), etc., provide photometric information of interesting point and extended sources, which require spectroscopic follow-up to understand the physical processes governing them in detail. 
VBT supports fast (F/3.25) and slow beam (F/13) observations, with the prime focus F/3.25 beam being particularly well-suited for fiber-fed multi-object and Integral field spectroscopy to study point and extended sources. To leverage this capability, a WFC system is designed to be installed at prime focus. This will help to improve the FoV and enable the mounting of several other back-end instruments to further science in the next decade. To reduce light loss, the design uses a minimal number of optical elements. The constraints on weight and volume at the prime focus have also led to the development of a compact optical system, making mechanical integration easier. A simple and effective three-element design with spherical lenses has been adopted, with careful selection of lens powers and spacing to control optical aberrations efficiently.

In this paper, we discuss in detail a compact and lightweight WFC system design for the VBT prime focus. We have designed and fabricated the WFC lens system having diameters smaller than 180~mm.  After its integration on a mechanical mount, the assembly has a total length of approximately 192~mm and a holder diameter of around 240~mm, with a total weight of about 15 kg. The fabrication was carried out at the optics and mechanical laboratory facility of the Indian Institute of Astrophysics (IIA).

\section{Optical design of wide-field corrector system}

 The WFC is designed for the VBT  to increase the FoV from 4$'$ to 30$'$ at its main focus.
 Table~\ref{tab:vbt_spec} outlines the VBT telescope parameters, including the primary mirror diameter, focal length, and f/ratio, which are fundamental for assessing the optical performance. The design of WFC carefully integrates atmospheric conditions—such as pressure, temperature, and humidity —to ensure its optimal performance under typical observing conditions at the VBT site. The atmospheric data collected by the VBO Weather Station during 2022 was used to calculate average values for atmospheric pressure, temperature, humidity, and other site-specific parameters, as shown in Table~\ref{tab:vbt_atm}. These atmospheric parameters are critical for the WFC design, as they influence calculations of atmospheric dispersion and its impact on the telescope's sensitivity. The seeing conditions at the VBT site vary between 1.5$''$ and 3.5$''$, with an average seeing of 2.5$''$ (\cite{2009MNRAS.395..593P}).

\begin{table}[htbp]
    \centering
    \caption{VBT Prime Focus Specifications}

    \label{tab:vbt_spec}
    \begin{tabular}{@{}l c@{}}
        \toprule
        \textbf{Parameter} & \textbf{Value} \\ 
        \midrule
        Primary Mirror Diameter & 2340~mm \\ 
        Focal Length & 7605~mm \\ 
        F/ratio & 3.25 \\ 
        Field of View (FoV) & $\approx$ 4$'$ \\ 
        Conic Constant & -1 \\ 
        Central Obscuration & 662~mm \\ 
        Image Scale & 27~arcsec/mm \\ 
        \bottomrule
    \end{tabular}
\end{table}

\begin{table}[htbp]
    \centering
    \caption{VBO Site Atmospheric Parameters}
    \label{tab:vbt_atm}
    \begin{tabular}{@{}l c@{}}
        \toprule
        \textbf{Atmospheric Parameters} & \textbf{Average Measurement} \\ 
        \midrule
        Air Temperature (K) & 296 \\ 
        Pressure (mbar) & 998.4 \\ 
        Humidity & 0.835 \\  
        Average Wind Speed (m/s) & 0.21 \\
        Altitude (m) & 725 \\ 
        Longitude & 78$^\circ49.6'$ E \\
        Latitude & 12$^\circ34.6'$ N \\
        \bottomrule
    \end{tabular}
\end{table}

\begin{figure}[htbp]
    \centering
 \includegraphics[width=0.8\linewidth]{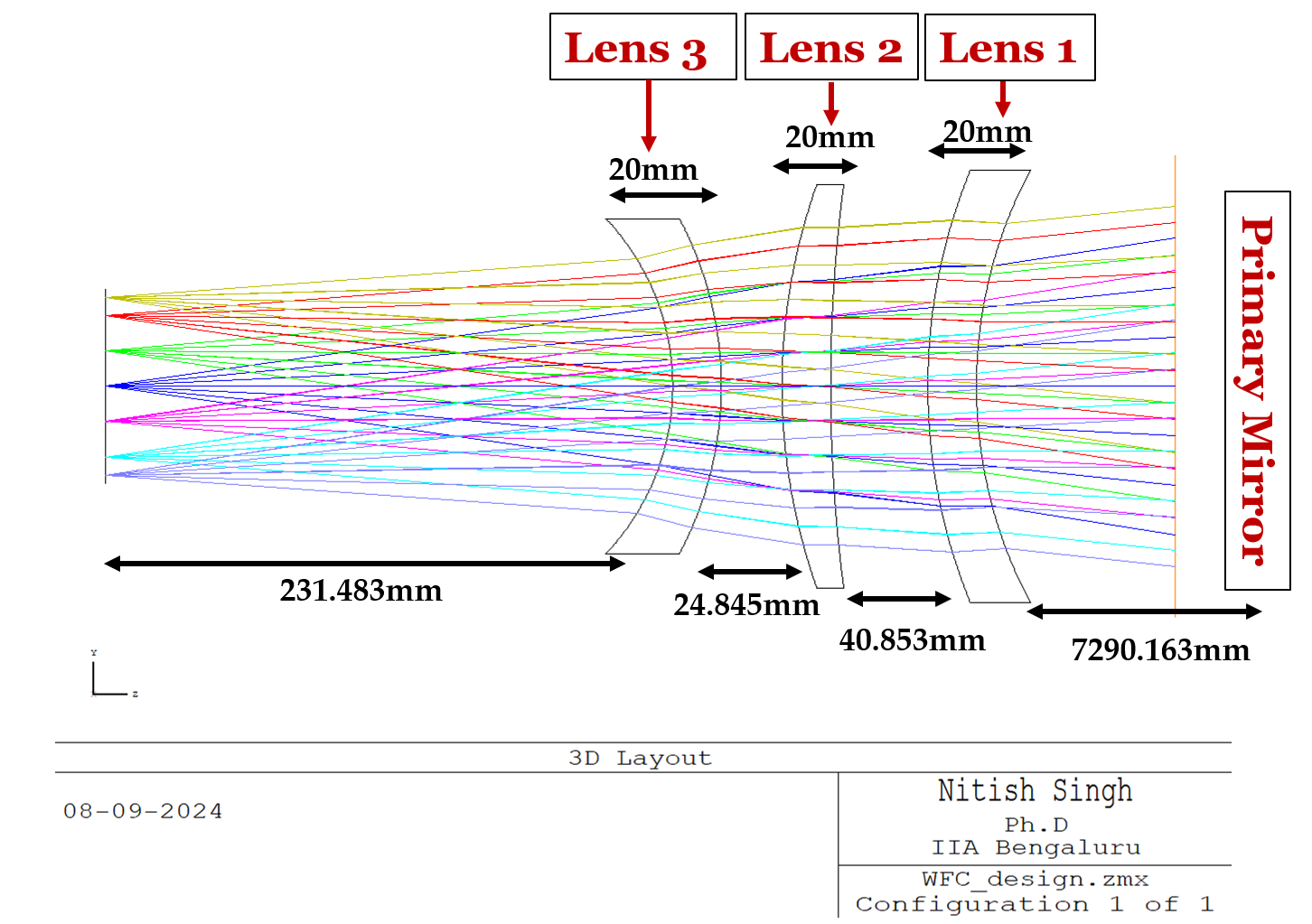}
    \caption{Optical layout of the WFC designed in ZEMAX for the VBT prime focus. }
    \label{fig:design_layout}

\end{figure}

 The WFC design was performed in ZEMAX, employing NBK7 glass. NBK7 was selected due to its favorable optical properties, including relatively low dispersion (with an Abbe number of ~64), which minimizes chromatic aberrations in optical systems. This material also exhibits high transparency in the visible spectrum (0.4~$\mu m$ to 0.9~$\mu m$), ensuring good light transmission, as detailed in the SCHOTT N-BK7 datasheet\footnote{\url{https://media.schott.com/api/public/content/41e799d0bf874807a0bb8e702fbb75b5?v=54856406}}. Additionally, NBK7 offers a balance between mechanical strength and thermal stability, making it well-suited for environmental conditions at VBO. Following the selection of NBK7, the WFC design converged after several iterations of optimization in ZEMAX, focusing on the radius of curvature (RoC), center thickness (CT), and inter-spacing distance between lenses to minimize aberrations and achieve the desired optical performance. The final configuration consists of three spherical lenses; Lens 1 and Lens 2 are negative lenses, and Lens 3 is positive, which can effectively corrects aberrations across the 30$'$ FoV. Each WFC lens has a CT of 20~mm, initially chosen under the assumption that the lens would require sufficient thickness to support the deeper curvatures for mechanical strength. The three lenses have diameters under 180~mm, as shown in Fig.~\ref{fig:design_layout}. Details of the optimized parameters, the RoC, CT, Clear Aperture Diameter (CA Dia), Edge Thickness (ET), and Material Thickness (MT), are presented in Table~\ref{tab:design_paramet}.

\begin{figure}[htbp]
    \centering
    \includegraphics[width=1\linewidth]{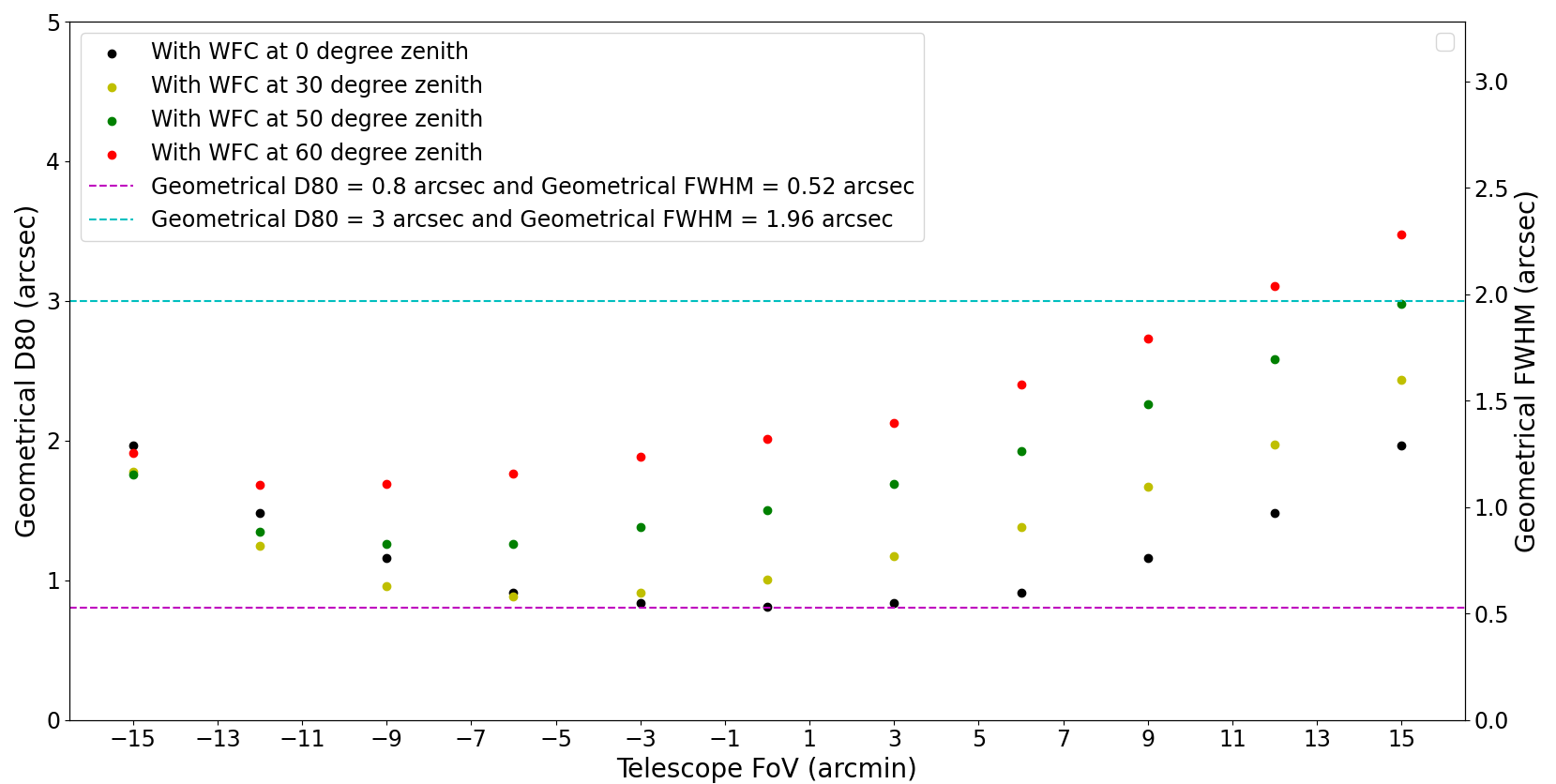}
    \caption{The figure presents the geometrical D80 performance of the WFC Across a 30$'$ FoV, at Zenith Angles from 0$^{\circ}$  to 60$^{\circ}$  and a Wavelength Range of 0.4 to 0.9~$\mu m$. }
    \label{fig:design_spot}
\end{figure}

 We incorporated the atmospheric parameters from Table~\ref{tab:vbt_atm} into the WFC design to analyze the Geometrical Encircled Energy Diameter (D80) and evaluate its performance across various zenith distances from 0$^{\circ}$ to 60$^{\circ}$ over a 30$'$ FoV. The mean geometrical D80 was found to be 1.22$''$ at zenith (0$^{\circ}$), increasing to 1.40$''$ at 30$^{\circ}$, 1.81$''$ at 50$^{\circ}$, and 2.25$''$ at a 60$^{\circ}$ zenith angle (see Fig.~\ref{fig:design_spot}). The Full Width at Half Maximum (FWHM) was then calculated using the relation given in \cite{2019sto..book.....T}.

\begin{equation}
\text{D80 = FWHM x 1.524}
\label{eq:net_seeing}
\end{equation}

The corresponding FWHM values are 0.8$''$ at zenith 0$^{\circ}$, 0.92$''$ at 30$^{\circ}$ zenith, 1.18$''$ at 50$^{\circ}$ zenith and 1.47$''$ at a 60$^{\circ}$ zenith angle (see Fig.~\ref{fig:design_spot}).  After introducing the WFC at the VBT prime focus, the telescope's f/ratio will change from 3.25 to 3.5, resulting in a modified plate scale from 37~$\mu m$/arcsec to 39.8~$\mu m$/arcsec.

In the optical design of the WFC, we introduced an Atmospheric Dispersion Compensator (ADC) to account for the effects of atmospheric dispersion, which can distort the quality of the images, especially at higher zenith angles. The ADC is composed of the same three lenses but Lens 1 and Lens 2 are movable, while Lens 3 remains fixed at the focal plane. This configuration allows for dynamic adjustment of the dispersion, ensuring minimal chromatic aberration across a wide range of zenith angles. With the introduction of the ADC into the WFC design and optimizing in Zemax, the system achieves an improved mean geometrical D80 of 1.33$''$ at 30$^{\circ}$ zenith, and the mean geometrical D80 of 1.96$''$ at a 60$^{\circ}$ zenith angle over a 30$'$ FoV.

After introducing atmospheric parameters from Table~\ref{tab:vbt_atm}, we observed that at 0$^{\circ}$ zenith, the WFC still performs symmetrically from -15$'$ to +15$'$ off-axis. But at higher zenith angles, like 30$^{\circ}$ and 60$^{\circ}$, the performance in D80 becomes asymmetric, with the range from 0$^{\circ}$ to +15$'$ performing worse than 0$^{\circ}$ to -15$'$ (see Fig.~\ref{fig:design_spot}). The reasons for this are ground-based telescopes are affected, significantly, by atmospheric dispersion, especially at non-zero zenith angles. As light from celestial objects transitions from the vacuum of space $n_{v}$ = 1 to Earth's atmosphere $ n_{a}$ = 1.0003, it undergoes refraction according to Snell's law (\cite{1999prop.book.....B}):

\begin{equation}
    n_{v} \sin(\theta_{v}) = n_{a} \sin (\theta_{a})
    \label{eq:atm_dispersion}
\end{equation}

At a zenith angle of 60$^{\circ}$ from the earth which is the incident angle ($\theta_{a}$), with an additional off-axis angular shift of $\pm$15$'$, the refraction angles ($\theta_{v}$) in vacuum differ slightly due to the nonlinear behavior of Snell’s law. The incident light first travels through the vacuum before entering Earth's atmosphere, where it undergoes refraction. The observed zenith angle from Earth is 60$^{\circ}$, and our FoV is 30$'$. At an off-axis position of +15$'$ from the center, the effective $\theta_{a}$ from Earth is 60.25$^{\circ}$, while at -15$'$, it is 59.75$^{\circ}$. To determine the corresponding vacuum $\theta_{v}$, we apply Equation~\ref{eq:atm_dispersion}. The calculations yield $\theta_{v}$ $\approx$ 59.78$^{\circ}$ for an off-axis shift of -15$'$ and $\theta_{v}$ $\approx$ 60.28$^{\circ}$ for +15$'$. This indicates that the light rays observed from Earth originate from slightly different angles in vacuum before being refracted by the atmosphere. As a result, the +15$'$ ray bends more than the -15$'$ ray, as refraction intensifies with steeper incidence angles. This differential bending introduces an asymmetry in optical performance, affecting metrics such as the D80. Specifically, deviations are more pronounced at +15$'$ than at -15$'$.

\section{Fabrication of Lenses and Quality Test}

\begin{figure}[htbp]
    \centering
    \includegraphics[width=1\linewidth]{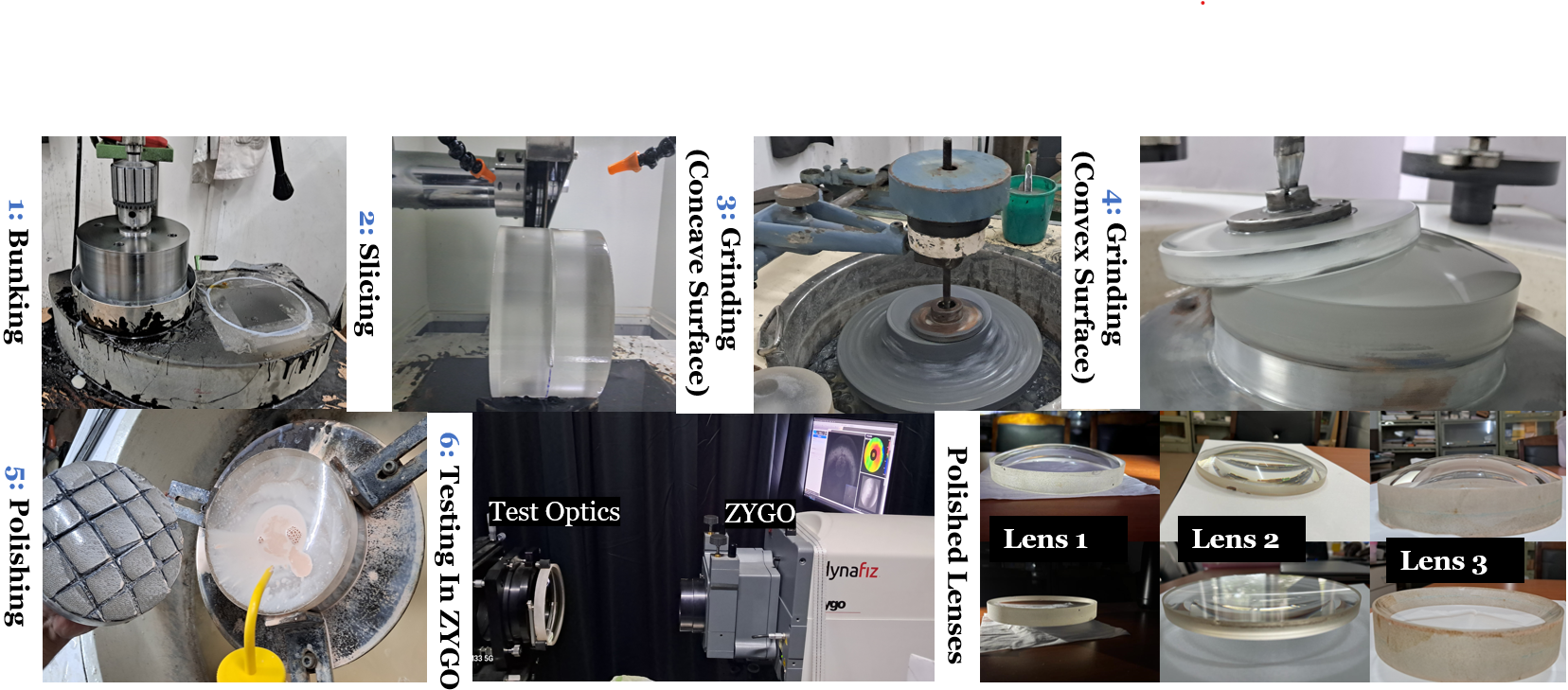}
    \caption{Fabrication and Testing of WFC Lenses: Key stages of the lens-making process including blank preparation, slicing, grinding, polishing, and surface testing using a ZYGO interferometer. }
    \label{fig:WFC_glass_making}

\end{figure}

Upon finalizing the design, the fabrication and testing of the WFC lenses were carried out at the IIA Optics Laboratory. The fabrication process started with a 400~mm diameter, 100~mm thick glass blank, which was cut into three pieces corresponding to diameters of 180~mm, 168~mm, and 140~mm, for Lens 1, Lens 2, and Lens 3, respectively (see Fig.~\ref{fig:design_layout}). These blanks were initially sliced to thicknesses slightly larger than the final required values of 42~mm, 26~mm, and 48~mm to allow for precision material removal during grinding. Coarse grinding was performed using $CaSO_{4}$ abrasives with grit sizes of 80, 120, 400, 600, 800, and 1000 to shape the lenses and remove surface imperfections. The sagitta of each lens surface was continuously monitored and adjusted using a spherometer to ensure that the desired curvatures were achieved within a tolerance limit of $\pm$1~mm in ROC, with a CT close to 20~mm. Once the required curvatures were attained, the lenses were polished using cerium oxide (Ce$_2$O$_3$) to achieve a surface figure accuracy of approximately 0.04~$\lambda$. Some key stages of the fabrication and testing process are shown in Fig.~\ref{fig:WFC_glass_making}.

After polishing, each lens underwent rigorous testing for Surface Figure Error (SFE) and RoC using ZYGO interferometer. The testing setup is illustrated in Fig.~\ref{fig:WFC_glass_making}. In this setup, the f/3 0.6328~$\mu m$ monochromatic beam from the ZYGO interferometer was directed onto the lens surface. A portion of the light reflected back to the ZYGO from the lens surface, and the resulting interference pattern was observed on a screen for SFE analysis. The SFE of each lens, measured using ZYGO interferometer, is provided in Table~\ref{tab:design_paramet}. Since light is refracted at both the entrance and exit surfaces of a lens, the contribution of a single surface to the overall wavefront error is approximately twice its SFE. To determine the net wavefront error of each lens, the individual errors from both surfaces were combined using the root sum square (RSS) method, a standard approach in optics for handling independent error contributions {\footnote{\url{https://wp.optics.arizona.edu/optomech/wp-content/uploads/sites/53/2016/10/S_Marshall.pdf?}}}. The wavefront error for each lens is given by Equation~\ref{eq:wave_error}.

\begin{equation}
\text{Wavefront error of Lens} =  \sqrt{(\text{concave SFE})^2 + (\text{convex SFE})^2}
\label{eq:wave_error}
\end{equation}

We achieved a RoC accuracy within $\pm1$~mm and a CT tolerance of $-2$~mm for all lenses, as outlined in Table~\ref{tab:design_paramet}. This table provides a detailed comparison of the design and manufactured values for each lens parameter, including CA Dia, RoC of Concave Surface (CCS), RoC of Convex Surface (CVS), CT, ET, MT, Root Mean Square (RMS) value of concave SFE, RMS value of convex SFE, and RMS value of wavefront error.

\begin{table}[htbp]
    \centering
    \caption{Optical Design and Post-Fabrication Parameters for WFC Lenses}
    \begin{tabular*}{\textwidth}{@{\extracolsep\fill}lcccccc}
        \toprule
        & \multicolumn{2}{@{}c@{}}{\textbf{LENS 1}} & \multicolumn{2}{@{}c@{}}{\textbf{LENS 2}} & \multicolumn{2}{@{}c@{}}{\textbf{LENS 3}} \\
        \cmidrule{2-3} \cmidrule{4-5} \cmidrule{6-7}
        Parameter & \textbf{Design} & \textbf{Fabricated} & \textbf{Design} & \textbf{Fabricated} & \textbf{Design} & \textbf{Fabricated} \\
        \midrule
        CA Dia (mm) & 178 & 178 & 166 & 166 & 138 & 138 \\
        RoC of CCS (mm) & 189.416 & 189.270 & 641.986 & 640.890 & 100.210 & 100.115 \\
        RoC of CVS (mm) & 234.794 & 235.160 & 246.999 & 247.180 & 144.553 & 143.422 \\
        CT (mm) & 20 & 18.6152 & 20 & 18.9705 & 20 & 18.475 \\
        ET (mm) & 25.142 & 22.513 & 10.927 & 10.705 & 30.247 & 28.42 \\
        MT (mm) & 41.595 & 37.286 & 25.242 & 23.961 & 47.442 & 46.18 \\
        RMS Concave SFE & 0.05~$\lambda$ & 0.026~$\lambda$ & 0.05~$\lambda$ & 0.029~$\lambda$  &0.05~$\lambda$ & 0.047~$\lambda$  \\
        RMS Convex SFE & 0.05~$\lambda$ & 0.019~$\lambda$   & 0.05~$\lambda$ & 0.027~$\lambda$  & 0.05~$\lambda$ & 0.037~$\lambda$     \\
        RMS Wavefront Error & 0.07~$\lambda$ & 0.032~$\lambda$ & 0.07~$\lambda$ & 0.039~$\lambda$ & 0.07~$\lambda$ & 0.06~$\lambda$ \\  
        \bottomrule
    \end{tabular*}
    \label{tab:design_paramet}
\end{table}

The tolerance requirements considered during the design and fabrication of the WFC are summarized in Table~\ref{tab:tolerances}. These tolerances include manufacturing tolerances, such as the permissible variations in the RoC of both convex and concave surfaces, CT, surface decenter, and surface irregularity Peak-to-Valley (PV), as well as alignment tolerances, including decenter in X, Y, and Z directions and allowable tilt about X and Y axes. These values were derived based on sensitivity analysis and the practical feasibility of fabrication and alignment. A compensating tolerance of $\pm5$~mm has been included for the Back Focal Length (BFL), considering cumulative deviations that may arise due to manufacturing and assembly variations. This tolerance framework ensured that the system performance remained within acceptable optical limits throughout fabrication and integration.
\begin{table}[htbp]
\centering
\caption{Manufacturing and Alignment Tolerances for the WFC optics.}
\begin{tabular*}{\textwidth}{@{\extracolsep\fill}lcccccc}
        \toprule

\textbf{Tolerance Type} & \textbf{Description} & \textbf{Lens 1} & \textbf{Lens 2} & \textbf{Lens 3} \\
\hline
\multirow{4}{*}{Manufacture } 
& RoC of CCS (mm) & $\pm 1$ & $\pm 1$ & $\pm 1$ \\
& RoC of CVS (mm) & $\pm 1$ & $\pm 1$ & $\pm 1$ \\
& Centre Thickness (mm) & $\pm 2$ & $\pm 2$ & $\pm 2$ \\
& Surface Decenter (arcmin) & $<1$ & $<1$ & $<1$ \\
& Surface Irregularity (PV) (0.6328~$\mu m$) & $<0.2\,\lambda$ & $<0.2\,\lambda$ & $<0.2\,\lambda$ \\
 \hline
\multirow{3}{*}{Alignment} 
& Decenter in X \& Y ($\mu$m) & $\pm 100$ & $\pm 100$ & $\pm 100$ \\
& Decenter in Z ($\mu$m) & $\pm 50$ & $\pm 50$ & $\pm 50$ \\
& Tilt about X \& Y (arcsec) & $\pm 60$ & $\pm 60$ & $\pm 60$ \\
 \hline
 \multirow{1}{*}{Compensator } 
& BFL (mm) & & & \multicolumn{2}{l}{$\pm$ 5} \\

        \bottomrule
\end{tabular*}
\label{tab:tolerances}
\end{table}

\section{D80 Performance Analysis of the Fabricated WFC}

Using the measured parameters of the individual lenses —thickness, RoC, and SFE (as presented in Table~\ref{tab:design_paramet}), we evaluated the average geometrical D80 performance over a 30$'$ FoV in ZEMAX. This analysis was conducted for a polychromatic wavelength range of 0.4 to 0.9~$\mu m$, considering different telescope pointing angles corresponding to zenith angles of 0$^{\circ}$, 30$^{\circ}$, 50$^{\circ}$, and 60$^{\circ}$. The distance between the lenses exhibited slight changes after incorporating the fabricated lens values. The effect of atmospheric dispersion as estimated from atmospheric parameters from Table~\ref{tab:vbt_atm} were incorporated into ZEMAX for geometrical D80 analysis. The mean geometrical D80 values for the WFC with the fabricated lenses were recalculated, yielding 1.36$''$ at zenith 0$^{\circ}$, 1.44$''$ at 30$^{\circ}$, 1.78$''$ at 50$^{\circ}$, and 2.21$''$ at a 60$^{\circ}$ zenith angle.  Similarly, the corresponding FWHM values were estimated, resulting in 0.89$''$ at zenith 0$^{\circ}$, 0.94$''$ at 30$^{\circ}$, 1.16$''$ at 50$^{\circ}$, and 1.45$''$ at 60$^{\circ}$ zenith angle. The calculated geometrical D80 values from ZEMAX were then combined with the VBT's atmospheric seeing disk. A seeing value of 2$''$ was chosen for the calculations as it is consistently observed at the site. We converted the seeing value to D80 using Equation~\ref{eq:net_seeing}. The net D80, accounting for both optical performance and atmospheric seeing, was determined using Equation~\ref{eq:net_D80}.

\begin{equation}
\text{Net D80} = \sqrt{\text{(1.524 x VBT seeing)}^2 + \text{(Geometrical D80)}^2} 
\label{eq:net_D80}
\end{equation}

\begin{figure}[htbp]
\centering
\includegraphics[width=1\linewidth]{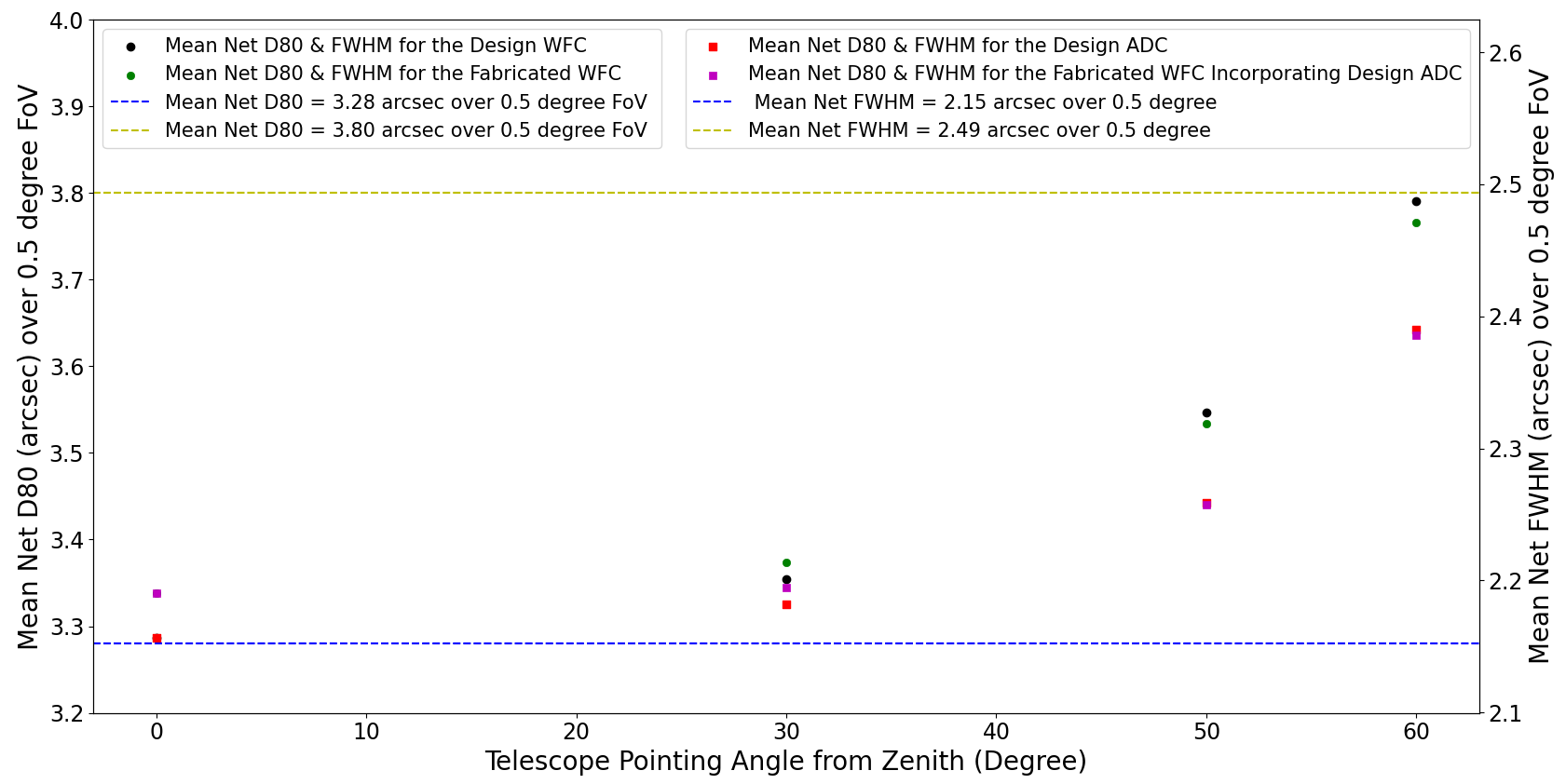} 
  \caption{Mean Net D80 of the Designed and Fabricated WFC over a 30$'$ FoV, considering VBT seeing conditions (2$''$). The calculation incorporates atmospheric parameters at different zenith angles, comparing the telescope's performance with the designed and fabricated WFC.}
   \label{fig:wfc_perform}
\end{figure}

This calculation accounts for the combined effects of the WFC’s geometrical performance and atmospheric seeing. The analysis was performed for both the designed and fabricated WFC. At a zenith angle of 0$^{\circ}$, the mean net D80 is 3.28$''$ for the designed WFC and 3.33$''$ for the fabricated WFC, with corresponding mean net FWHM values of 2.15$''$ and 2.18$''$. At 30$^{\circ}$ zenith, the mean mean net D80 is 3.35$''$ and 3.37$''$, with mean net FWHM values of 2.19$''$ and 2.21$''$. At 60$^{\circ}$ zenith, the mean net D80 is 3.79$''$ for the designed WFC and 3.76$''$ for the fabricated WFC, with corresponding mean net FWHM values of 2.48$''$ and 2.46$''$ (see Fig.~\ref{fig:wfc_perform}).

In Fig.~\ref{fig:wfc_perform}, it is evident that for a 30$'$ FoV and zenith angles ranging from 0$^{\circ}$ to 60$^{\circ}$, the mean net D80 and mean net FWHM performance remain better than 3.80$''$ and 2.49$''$, respectively, after accounting for the 2$''$ VBT site seeing. After introducing the ADC, we achieved an improvement of approximately 0.1$''$ in both net D80 and FWHM performance. The simulations were performed to assess the image quality with and without ADC integration in WFC. This comparison further confirms that the fabricated WFC closely aligns with the designed performance, both in optical parameters and overall image quality. Furthermore, the f/ratio of the telescope, which is currently 3.25, increases to 3.48 after incorporating the fabricated WFC. The VBT prime focus plate scale changes from 37~$\mu m$ to 39.6~$\mu m$.

\section{Mechanical Design, Fabrication, and Quality Check of the WFC Unit}

\begin{figure}[htbp]
    \centering
    \includegraphics[width=0.8\linewidth]{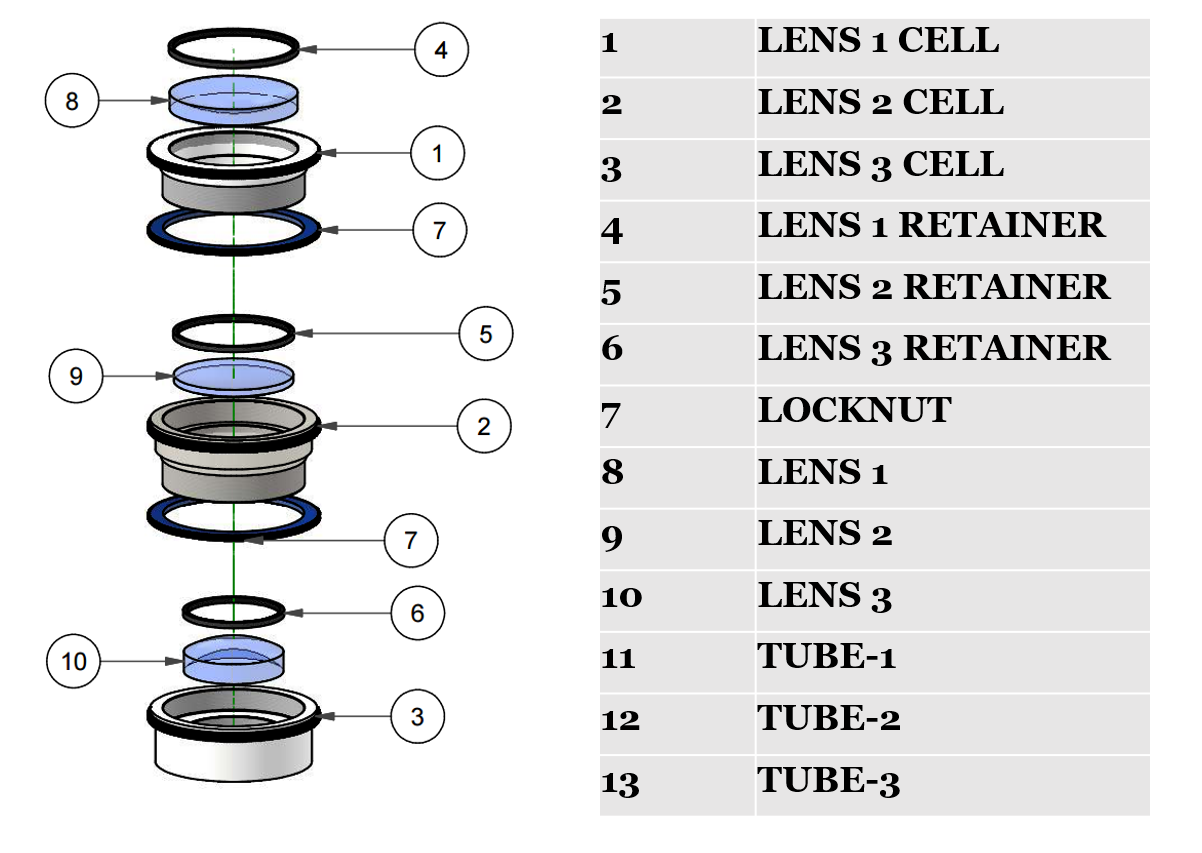}
    \caption{WFC Lens Assembly Holder}
    \label{fig:wfc_mech_part}

\end{figure}

The mechanical design of the WFC mount was developed using Autodesk Inventor, allowing for the creation of an accurate and structurally reliable assembly to securely hold the optical components, as shown in Fig.~\ref{fig:wfc_mech_part}, the design prioritizes precision, stability, and ease of assembly while maintaining minimal mechanical stress on the lenses. Aluminum 6061 was selected as the primary material for the WFC mount due to its advantageous properties, including low density (2.7 $g/cm^3$), high strength-to-weight ratio, excellent corrosion resistance, ease of machinability, widespread availability, and cost-effectiveness. The mount features a lens cell designed to, securely hold the lenses, a retainer ring controls the position of the lens and the locknut maintains the relative position of the lenses for obtaining precise optical alignment. To further safeguard the integrity of the lenses, flexible supports were incorporated to distribute mechanical stresses evenly, reducing potential risks of deformation or misalignment. Post-fabrication, the unit underwent a surface treatment process involving dull matte black anodization to minimize stray light leakage. At present, the ADC is not integrated into the current mechanical design of the WFC mount. For alignment and verification, a suite of precision metrology tools was used, including dial gauges, FARO gauge arms, laser alignment systems, theodolite, and interferometric setups. These techniques were guided by alignment strategies similar to those employed in the 4MOST WFC assembly \cite{2022SPIE12184E..6VC}, providing a reliable benchmark for our procedures. In the preliminary testing phase, the lenses were temporarily held using mechanical locknuts instead of RTV pads, a choice made to facilitate easier re-adjustments during iterative alignment trials. These procedures were instrumental in verifying optical axis alignment, minimizing decenter and tilt errors, and ensuring mechanical stability before full integration. A summary of the manufacturing and alignment tolerances is provided in Table~\ref{tab:tolerances}.

\begin{figure}[htbp]
  \centering
  \begin{subfigure}[b]{0.5\textwidth} 
    \centering
    \includegraphics[width=\textwidth]{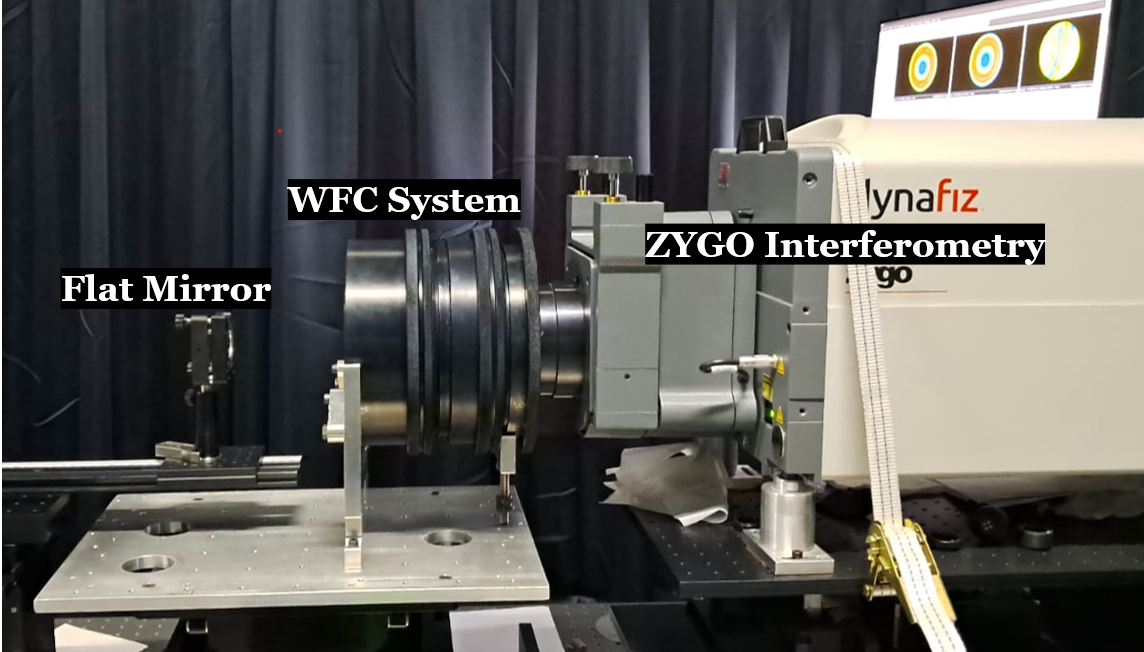}
    \caption{}
    \label{fig:zygo_wfc}
  \end{subfigure}
  \hfill
  \begin{subfigure}[b]{0.48\textwidth} 
    \centering
    \includegraphics[width=\textwidth]{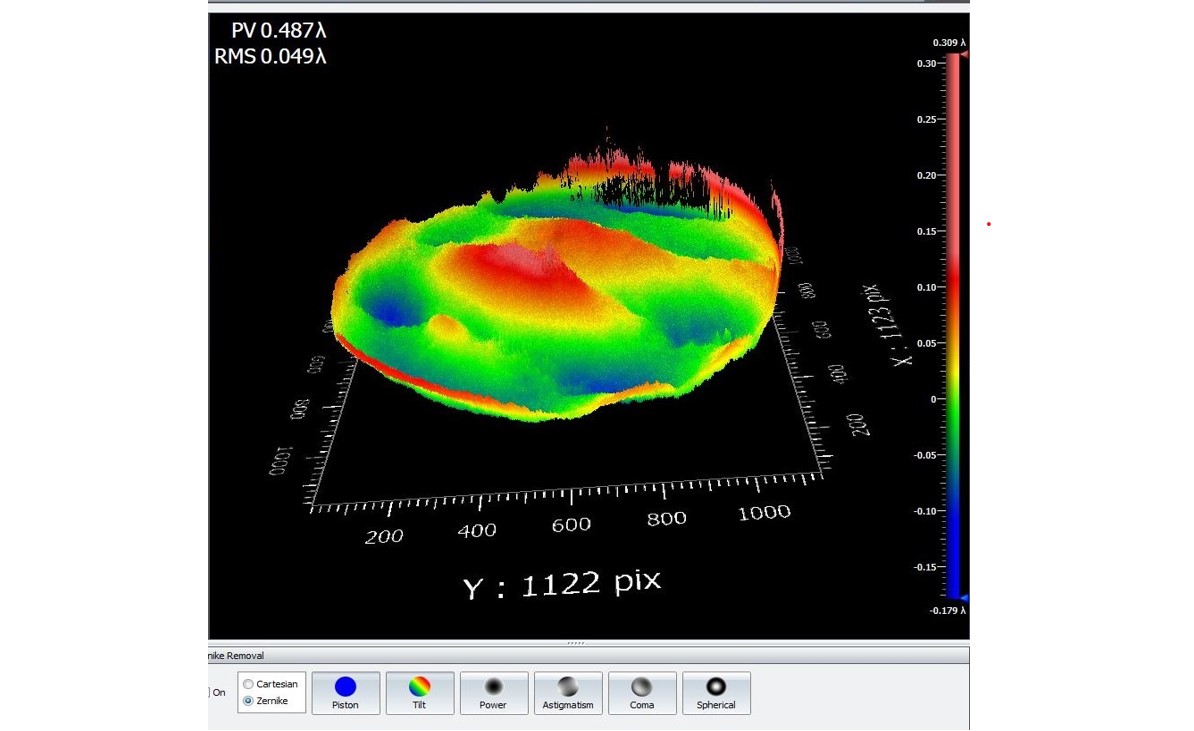} 
    \caption{}
    \label{fig:wfc_figure}
  \end{subfigure}
  \caption{(a) ZYGO interferometer setup used to evaluate the WFC unit after fabrication. This test provided data on the  RoC and quantified the SFE introduced by the WFC.
(b) Wavefront error map (0.05~$\lambda$) of the complete WFC system, illustrating the lens surface quality and its impact on overall optical performance. }
  \label{fig:wfc_zygotesting}
\end{figure}

After assembling all the lenses in the mechanical mount, the wavefront accuracy of the WFC was evaluated using ZYGO interferometry, as illustrated in Fig.~\ref{fig:zygo_wfc}. In this setup, an f/3 beam at 0.6328~$\mu m$ wavelength passes through the WFC, reflects off a flat mirror, and then traverses back through the WFC before interfering within the ZYGO interferometer. This double-pass configuration effectively measures the wavefront error introduced by the WFC. The measured wavefront error was found to be 0.05~$\lambda$, as shown in Fig.~\ref{fig:wfc_figure}.

\section{Testing and Validation of the WFC in Lab Setup}

After confirming that the WFC parameters (like RoC, CT, and wavefront error) met the tolerance limits, laboratory testing was conducted to validate both on-axis and off-axis performance. Prior to physical testing, an optical setup was designed in ZEMAX using the fabricated WFC parameters for each lens (as listed in Table~\ref{tab:design_paramet}). The setup included three achromatic doublets obtained from Edmund Optics \footnote{\url{https://www.edmundoptics.com}}, with specifications detailed in Appendix~\ref{appendix:lab_lens_specs} . The D80 performance without the WFC ($D80_{\text{NoWFC,design}}$), was calculated in ZEMAX for the on-axis beam and found to be 28.04~$\mu m$. This represents the optimal performance of the achromatic doublet setup only. The WFC was then integrated using the fabricated lens parameters (RoC, CT, and wavefront error). To simulate off-axis performance, the WFC was tilted and decentered in ZEMAX. This approach avoids introducing any off-axis aberrations caused by the test lens, ensuring that only the WFC’s off-axis performance is evaluated. Using the full system D80 performance ($D80_{\text{full,design}}$), the WFC-only contribution ($D80_{\text{WFC,only,design}}$), were calculated using:

\begin{equation}
D80_{\text{WFC,only,design}} = \sqrt{D80_{\text{full,design}}^2 - D80_{\text{NoWFC,design}}^2}
\label{eq:WFC_only_design}
\end{equation}

\begin{figure}[htbp]
  \centering
    \includegraphics[width=0.85\linewidth]{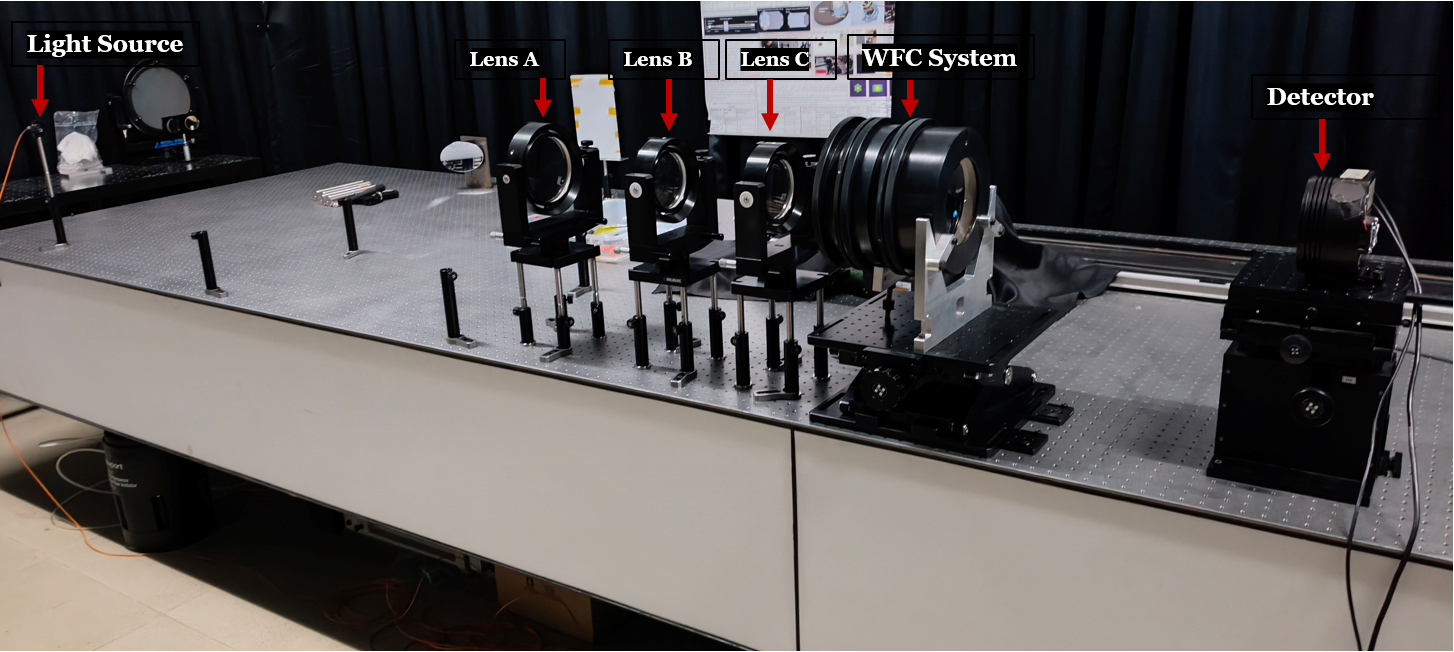}
  \caption{Laboratory setup for testing the WFC using off-the-shelf lenses from Edmund Optics}
  \label{fig:lab_testing_wfc}
\end{figure}

Detailed D80 values are provided in Appendix~\ref{appendix:D80_WFC_design}. ZEMAX simulations was performed restricting to 0.4-0.7~$\mu m$ wavelength range to work in the allowable range of the test lens. The laboratory experimental setup was assembled as shown in Fig.~\ref{fig:lab_testing_wfc}. The system was illuminated using a white light source coupled to a 100 $\mu$m core diameter f/3 fiber, and the resulting spot size was captured with a CMOS detector having a pixel scale of 20~$\mu m$/pixel. 

\begin{figure}[htbp]
    \centering
    \includegraphics[width=0.6\linewidth]{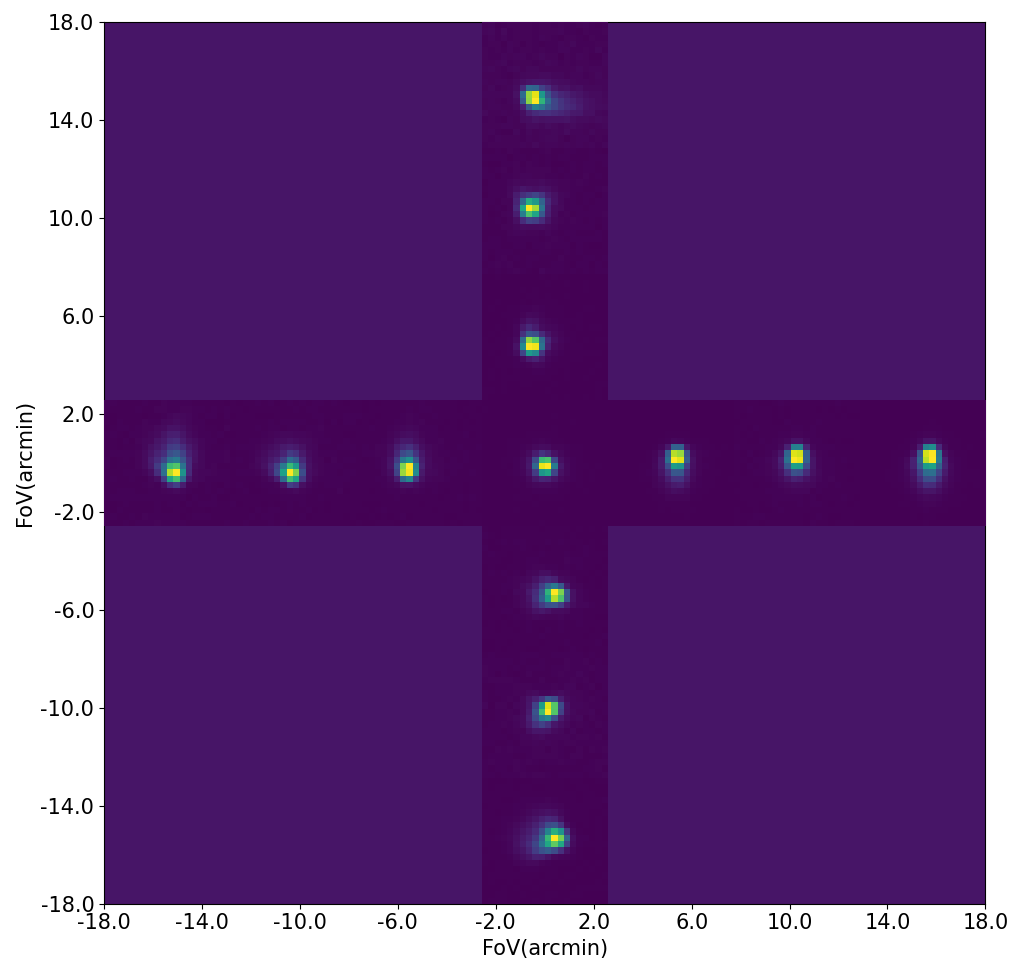}
    \caption{Off-axis performance of the WFC in the x and y directions. The measured spot size is 2.41~pixel for the on-axis beam and 2.94~pixel for the extreme field.}
    \label{fig:lab_xydirs_perf}
\end{figure}

Performance tests for on-axis and off-axis configurations were conducted with field angles from -15$'$ to 15$'$ in both x and y directions. Results showed an on-axis spot size of 48.2~$\mu m$ (2.41~pixel), increasing to 58.8~$\mu m$ (2.94~pixel) at off-axis (detailed values in Appendix~\ref{tab:D80_labWFC_appendix}), as shown in Fig.~\ref{fig:lab_xydirs_perf}.

After obtaining the images from the full setup, we calculated the Gaussian FWHM for each spot using IRAF \citep{1986SPIE..627..733T}. The FWHM values were then converted to D80 diameters using Equation~\ref{eq:net_seeing}. To quantify the D80 introduced solely by the WFC, we used the following equation:

\begin{equation}
D80_{\text{WFC only,lab,x}} = \sqrt{D80_{\text{full,lab,x}}^2 - D80_{\text{NoWFC,design}}^2}
\label{eq:WFC_onl_labx}
\end{equation}

\begin{equation}
D80_{\text{WFC only,lab,y}} = \sqrt{D80_{\text{full,lab,y}}^2 - D80_{\text{NoWFC,design}}^2}
\label{eq:WFC_onl_laby}
\end{equation}

\begin{figure}[htbp]
    \centering
    \includegraphics[width=0.8\linewidth]{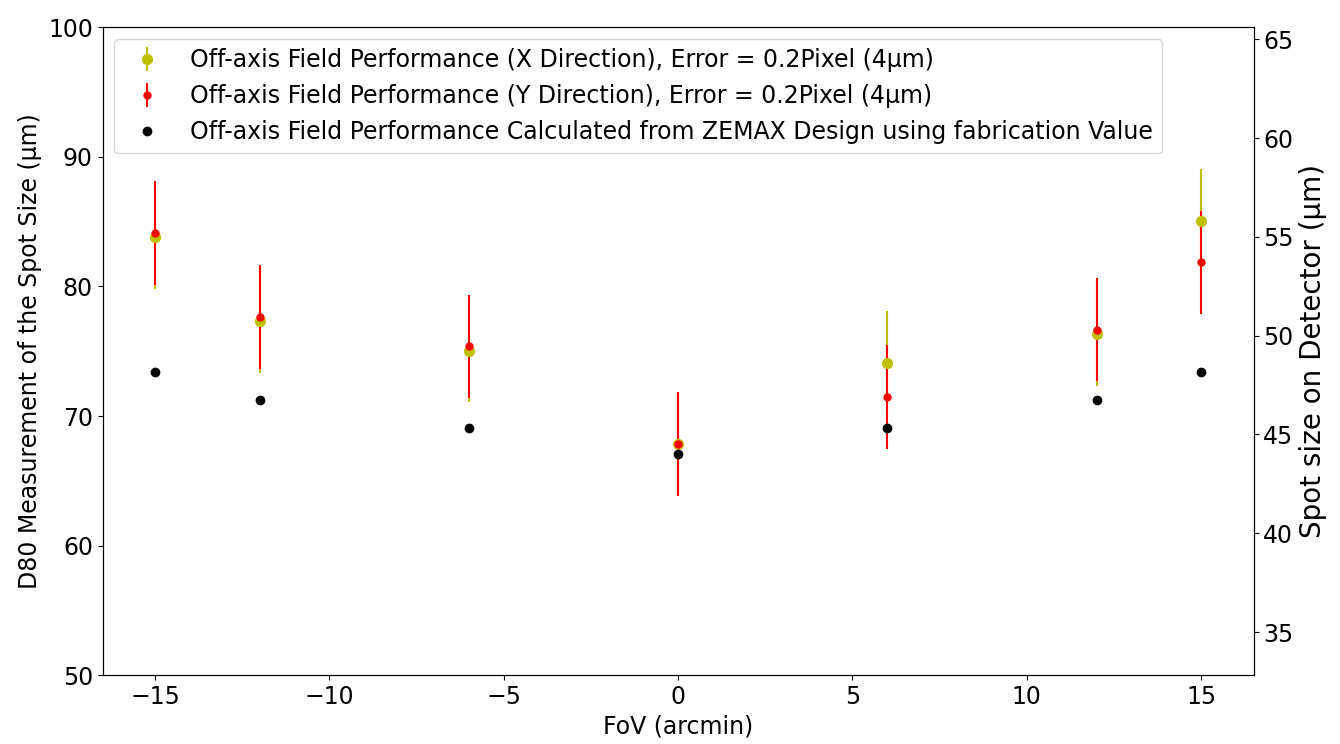}
    \caption{Comparison of design and laboratory D80 performance across off-axis angles with the WFC tilted and decentered in both the x and y directions. The lab performance of the WFC deviates by approximately 0.1~pixel for the on-axis and 0.5~pixel for the off-axis from the design value. }
    \label{fig:lab_offaxis}
\end{figure}

Where $D80_{\text{WFC only,lab,x}}$ and $D80_{\text{WFC only,lab,y}}$ represent the D80 values introduced solely by the WFC at different off axis angles along the x and y directions, respectively (detailed values are provided in Appendix~\ref{tab:D80_labWFC_appendix}). We then plotted the D80 values at different off-axis angles, ranging from -15$'$ to +15$'$, for both the design and the actual laboratory setup (see Fig.~\ref{fig:lab_offaxis}). The Comparison between the D80 introduced solely by the WFC in the design setup ($D80_{\text{WFC,only,design}}$) and the laboratory setup ($D80_{\text{WFC,only,lab}}$) showed a deviation of approximately 2~$\mu m$ (0.1~pixel) for the on-axis beam and around 10~$\mu m$ (0.5~pixel) for the extreme off-axis case. The results show that the D80 values introduced by the WFC in the lab closely match those predicted by the design. The error bars represent measurement uncertainties due to tilting and decentering of the WFC in the x and y directions during laboratory testing. These uncertainties, calculated using the standard deviation method, were approximately 0.2~pixel (4~$\mu m$). These results confirm the WFC's reliable performance across the FoV.

\section{Transmission Efficiency Calculation of WFC}

\begin{figure}[htbp]
  \centering

    \includegraphics[width=0.7\linewidth]{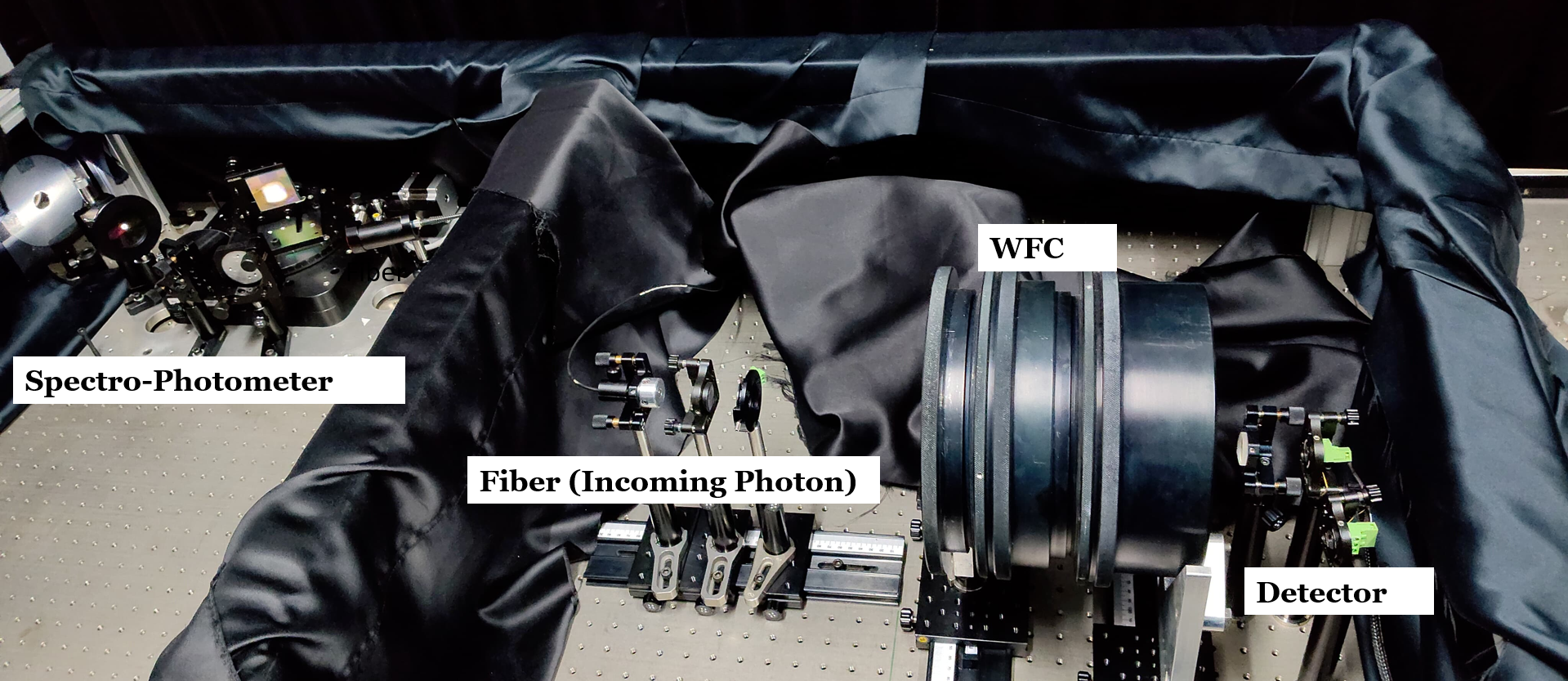}
\caption{ Experimental setup for measuring the transmission efficiency of the WFC using a spectrophotometer.}
 \label{fig:effi_lab}
\end{figure}

We evaluated the transmission efficiency of WFC system using a spectrophotometer operating in the wavelength range from 0.4~$\mu m$ to 0.75~$\mu m$, along with a lab CMOS detector to capture the spot size. The laboratory setup is shown in Fig.~\ref{fig:effi_lab}. The WFC system consists of three lenses, which means six optical surfaces. The transmission efficiency (T) of WFC is calculated as described in \cite{1999prop.book.....B}:

\[
T = \left(1 - R \right)^6
\]

where (R) is the reflectivity at each surface, which is determined as following,

\[
R = \left(\frac{n_1 - n_2}{n_1 + n_2} \right)^2
\]

where $n_{2}$ is the refractive index of air ($n_{2}$ = 1), and $n_{1}$  is the refractive index of NBK7 glass, which varies with wavelength. 

\begin{figure}[htbp]
  \centering
    \includegraphics[width=1\linewidth]{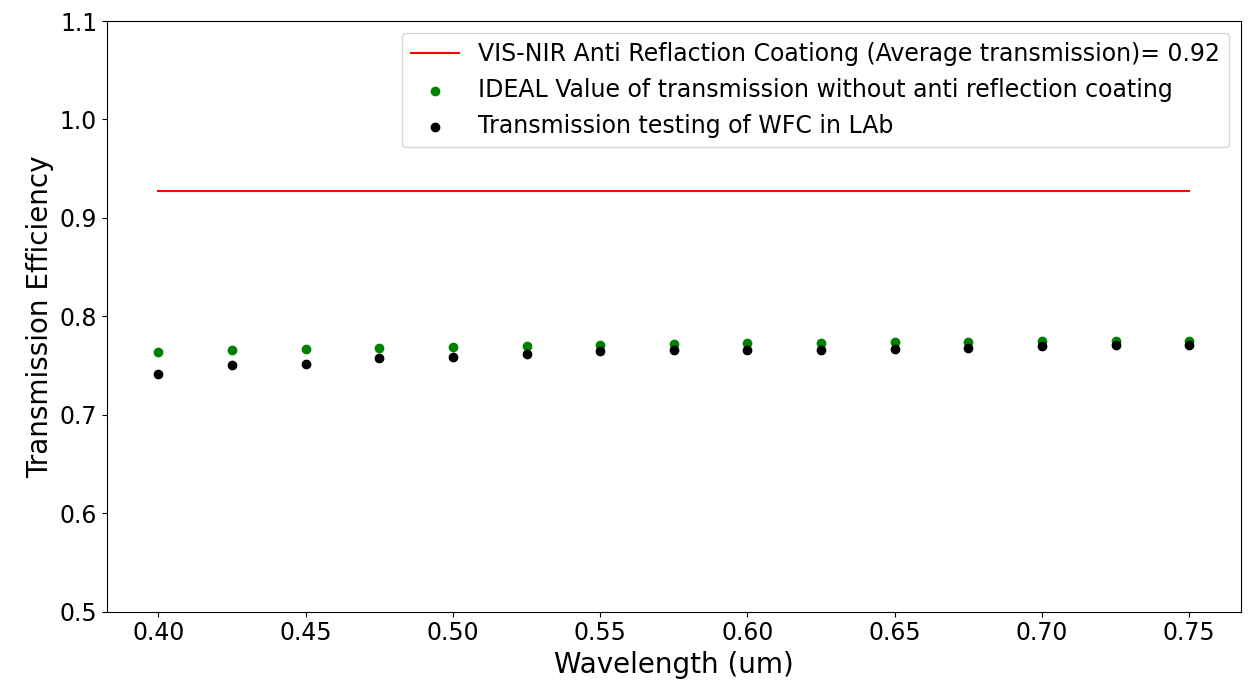}

\caption{Wavelength versus transmission efficiency plot, comparing the ideal transmission efficiency of the WFC (calculated using refractive indices from the SCHOTT N-BK7 datasheet) with lab test results in the 0.4 to 0.75~$\mu m$ wavelength range. The plot also indicates the transmission efficiency of the VIS-NIR coating from Edmund Optics for NBK7 glass.}
\label{fig:eff_plot}
\end{figure}

Using the \textit{SCHOTT N-BK7 datasheet}\footnote{\url{https://media.schott.com/api/public/content/41e799d0bf874807a0bb8e702fbb75b5?v=54856406}}, we calculated the refractive indices at different wavelengths to estimate the ideal transmission efficiency. We then measured the actual transmission of the WFC using a spectrophotometer (see Fig.~\ref{fig:effi_lab}). At 0.4~$\mu m$, the estimated transmission was $\approx$76.3\%, while the measured value was $\approx$74.1\%. Similarly, at 0.575~$\mu m$, the ideal transmission was $\approx$77\%, and the measured value was $\approx$76.1\%. At 0.75~$\mu m$, both the estimated and measured transmissions were nearly identical at $\approx$77.5\% and $\approx$77\%, respectively, showing minimal deviation at longer wavelengths (see Fig.~\ref{fig:eff_plot}). Overall, the measured transmission was on average 1.4\% lower than the theoretical predictions over the 0.4–0.75~$\mu m$ range, which aligns with the operational limitations of our spectrophotometer’s lamp. To further enhance transmission efficiency, we plan to apply a broadband anti-reflection coating. The \textit{VIS-NIR coating from Edmund Optics} \footnote{\url{https://www.edmundoptics.in/knowledge-center/application-notes/lasers/anti-reflection-coatings/?srsltid=AfmBOorZN9cdFAvqJLOQyN_38fGu1ipANpjY7KC_hZD_PfhEqZVW44Yj}}, with specifications of $T_{\text{avg}} \leq 98.75\%$ ($R_{\text{avg}} \leq 1.25\%$)  across 0.4–1.0~$\mu m$, is expected to improve the overall transmission performance.

\section{SUMMARY AND DISCUSSION}

We have designed, fabricated, and assembled a three-element WFC system for the VBT. The WFC system, consisting of three spherical lenses to cover a polychromatic wavelength range, will enhance the corrected FoV from 4$'$ to 30$'$ at prime focus of VBT. The aberration correction, tolerances, and atmospheric dispersion analysis of WFC system are performed using ZEMAX to ensure its precision and stability within defined tolerance limits. The fabricated lenses met stringent tolerance limit, with a $\pm1$~mm deviation in the RoC and $\pm2$~mm deviation in CT compared to the design values (see Table~\ref{tab:design_paramet}). 
A mechanical mount was designed and developed for the WFC, onto which the lenses were integrated. The wavefront error was then measured using ZYGO interferometer and the wavefront error is 0.05~$\lambda$.
The transmission efficiency of the WFC system was measured using a spectrophotometer that showed 74 to 77\% over the 0.4–0.75~$\mu m$ range,  agreeing with the theoretical calculations 76 to 77\%.  The transmission is expected to increase to 92\%  after applying a VIS-NIR anti-reflection coating to the lens system. 
A dedicated laboratory setup was then constructed to evaluate the WFC's performance over an angular range of -15$'$ to +15$'$,  using achromatic lenses and a polychromatic light source delivered through a 100 $\mu$m core diameter f/3 fiber. Initially designed in ZEMAX and later implemented in the lab, the actual spot size deviated from the design by only approximately 0.1~pixel for the on-axis case and 0.5~pixel for the extreme off-axis field. 

The D80 performance of the WFC is a key measure of its image quality. The D80 measured from WFC in the laboratory setup across the FoV was found to be in good agreement with the design predictions. 
The mean geometrical D80 for the designed WFC, as obtained from ZEMAX simulations, was 1.22$''$ (48.31~$\mu m$) at zenith (0$^\circ$), increasing to 1.40$''$ at 30$^\circ$ zenith angle and 2.25$''$ at 60$^\circ$.  Taking into account the 2$''$ seeing conditions at VBT, the average net seeing was determined to be 2.19$''$ at zenith (0$^{\circ}$), 2.21$''$ at 30$^{\circ}$ zenith, and increased to 2.47$''$ at a 60$^{\circ}$ zenith angle over the same 30$'$ FoV. These values were recalculated using the optical parameters of the fabricated WFC, showing similar D80 performance (see Fig.~\ref{fig:wfc_perform}).

We compare the perfomance of our three-spherical lens WFC system designed for VBT with similar systems on different telescopes across the world.   
The 2.2 m UHWFI achieves a geometrical D80 of 0.69$''$ (or 45.72~$\mu m$) over 0.4–1.0~$\mu m$ \cite{2006PASP..118..780H}. The 4.01m Mayall Telescope provides a 1.5$^\circ$ corrected FoV over the 0.4–1.0~$\mu m$ range, achieving a geometrical D80 of 0.5$''$ at zenith (0$^\circ$) and 1.5$''$ at 60$^\circ$ \cite{2024AJ....168...95M, 2014SPIE.9151E..1MS}. Similarly, the 3.9m AAT features a WFC with a 2.1$^\circ$ FoV, achieving D80 values of 0.5$''$ at zenith and 1.5$''$ at 60$^\circ$ \cite{1994ApOpt..33.7362J}. The CFHT demonstrates superior performance, achieving a D80 of 0.3$''$ at zenith over a broad wavelength range of 0.35–2.0~$\mu m$. 

Our compact and lightweight WFC design demonstrated promising results considering the spherical lenses in the system. However, aspherical lens systems show better performance over spherical lens systems. 
At present as a first step to improving the FoV, our design incorporates only spherical lenses, that were fabricated in the IIA laboratory, where a polishing facility for spherical lenses is readily available. A design incorporating aspherical lenses is underway, and a design for this is in progress. This design will also include an ADC in the mechanical setup, and we plan to integrate it in the near future. 

The WFC will be installed at the prime focus of the VBT. The mechenical mount for WFC system to install at VBT prime focus is already designed and is being fabricated.  
The enhanced FoV of 30$'$ (70mm in size) after introducing the WFC at VBT prime focus enables simultaneous use of two spectrographs, the HRES and the OMRS, feeding light through fibers from the prime focus. 
The availability of two instruments in same night will enhance the VBT observing capability to two-fold. 
The fiber-fed setup for OMRS and HRES at VBT prime focus are already designed, and the details are available at \cite{2024arXiv240615591S}. The integral field unit in the above set-up will allow astronomers, additionally, to obtain spectra of extended objects like planetary nebulae, galaxies, etc. The wide field (70mm) can also accommodate 40 fibers operated using the Alpha-Beta positioning system the design for which is under progress. All these multiple arrangements will allow simultaneous observation of multiple objects across the 30$'$ FoV, further expanding the VBT's capability for multi-object spectroscopy.

\bmhead{Acknowledgements}
 
We sincerely thank the VBT staff for sharing information about the telescope, Dr. S Muneer, and the VBO weather station team for providing weather data.  Special thanks to Remya B. S., and Debadutta for assisting in the testing of the WFC and   Prasobh P. and Ananda for helping with mechanical fabrication. Finally, we thank IIA for funding this project.


\section*{Declarations}

\subsection{Availability of Data and Materials} \label{material_availibility}  

The data supporting the findings of this study are available from the corresponding author, Nitish Singh, upon reasonable request. The following software tools were used in our analysis: IRAF \citep{1986SPIE..627..733T}, Numpy \citep{harris2020array}, Astropy \citep{2022ApJ...935..167A}, Scipy \citep{2020SciPy-NMeth}, Matplotlib \citep{Hunter:2007}, ZEMAX \citep{Zemax}, Autodesk Inventor \citep{Autodesk_Inventor}, and AutoCAD \citep{Autodesk_AutoCAD}.

\subsection{Funding declaration}
This research was made possible through financial support from the Indian Institute of Astrophysics (IIA) under the Department of Science and Technology (DST) in India.

\subsection{Author contribution declaration}

YBK, SSR, and RS conceptualized the project. NS and SSR conducted the optical design and testing of the WFC lens system. NS and RS performed the data analysis. NS, PMMK, KS, and SD contributed to the mechanical mount design. NS, RB, GN, and CC fabricated the WFC lenses, while SK and FXRJ fabricated the mechanical mount. NS, YBK, RS, and SSR contributed to the writing of the manuscript. All authors reviewed and approved the final manuscript.


\subsection{Declaration of competing interest}

The authors declare no conflict of interest with any known people.

\begin{appendices}

\section{Laboratory Testing and D80 Performance Analysis}

\subsection{Laboratory Test Setup}

The lenses used in our laboratory to evaluate the WFC performance, both on and off-axis, are sourced from Edmund Optics (Lens A: Model 54569; Lenses B and C: Model 54568). These lenses are essential for assessing the optical quality and aberrations introduced by the WFC in our test setup. The working range of these lenses is 0.4-0.7~$\mu m$. The following table provides the specifications of the lenses used in the laboratory tests:
\label{appendix:lab_lens_specs}
\begin{table}[h]
\caption{Laboratory Lens Details}\label{tab:lab_lens_spe_appendix}
\begin{tabular}{@{}lllll@{}}
\toprule
\textbf{Parameters} & \textbf{Lens A} & \textbf{Lens B} & \textbf{Lens C}   \\
\midrule
Diameter (mm) & 140 & 116 & 116   \\
Effective Focal Length (EFL) (mm) & 1900.24 $\pm$ 1  & 1524.73 $\pm$ 1  & 1524.73 $\pm$ 1 &  \\
Back Focal Length (BFL) (mm) & 1887.58 $\pm$ 1  & 1514.53 $\pm$ 1  & 1514.53 $\pm$ 1 &  \\    
\botrule
\end{tabular}
\end{table}

\subsection{Additional D80 Performance Details}  

Here, we present the D80 values for the full system (with the  WFC) and the D80 values introduced by the WFC alone, as calculated for the telescope setup. These values take into account the tilt and decenter effects across various FoV angles, providing a detailed performance analysis for different positions within the FoV. 
\label{appendix:D80_WFC_design}

\begin{table}[h]
\caption{D80 values for the full system (with the WFC) and introduced by the WFC alone}
\label{tab:D80_WFC_design_appendix}
\begin{tabular}{@{}lccccccc@{}}
\toprule
\textbf{FoV (degree)} & \textbf{-0.25} & \textbf{-0.2} & \textbf{-0.1} & \textbf{0} & \textbf{0.1} & \textbf{0.2} & \textbf{0.25}  \\
\midrule
$D80_{\text{full,design}} (\mu m)$ & 78.6 & 76.62 & 74.62 & 72.76 & 74.62 & 76.62 & 78.6  \\  
$D80_{\text{WFC only,design}} (\mu m)$ & 73.37 & 71.25 & 69.09 & 67.08 & 69.09 & 71.25 & 73.37   \\  
\botrule
\end{tabular}
\end{table}

This table highlights the performance improvements introduced by the WFC, where the full system (with the WFC) shows an increase in D80, especially at the edge of the FoV. Meanwhile, the WFC alone demonstrates the intrinsic performance of the optical design, achieving minimal aberrations at the center of the field.

\subsection{D80 Performance in the Laboratory Setup}  

In this section, we present the D80 values obtained through laboratory testing for both the full system (with the WFC) and the WFC alone. The values are derived from the measured spot sizes, which are then converted into D80 using the formula from Equation~\ref{eq:net_seeing}. This table provides the detailed D80 values in both the x and y directions for various FoV angles.
\label{appendix:D80_lab_values}

\begin{table}[h]
\caption{D80 values for the full system (with the WFC) and introduced by the WFC alone in the lab setup}
\label{tab:D80_labWFC_appendix}
\begin{tabular}{@{}lccccccc@{}}
\toprule
\textbf{FoV (Degree)} & \textbf{-0.25} & \textbf{-0.2} & \textbf{-0.1} & \textbf{0} & \textbf{0.1} & \textbf{0.2} & \textbf{0.25}  \\
\midrule
$D80_{\text{full, lab,x}} (\mu m)$ & 88.39 & 82.29 & 80.16 & 73.45 & 79.25 & 81.38 & 89.61  \\  
$D80_{\text{full, lab,y}} (\mu m)$ & 88.69 & 82.60 & 80.46 & 73.45 & 76.81 & 81.68 & 86.56  \\  
$D80_{\text{WFC only,lab,x}} (\mu m)$ & 83.78 & 77.33 & 75.04 & 67.84 & 74.06 & 76.34 & 85.06   \\  
$D80_{\text{WFC only,lab,y}} (\mu m)$ & 84.10 & 77.64 & 75.37 & 67.84 & 71.45 & 76.67 & 81.84   \\  
\botrule
\end{tabular}
\end{table}

This table provides valuable insight into the WFC's performance in the lab, showing the variation in D80 with respect to the FoV angles. The WFC alone demonstrates significant improvement in D80, especially in the center of the FoV, highlighting its effectiveness in reducing aberrations.

\end{appendices}

\input{sn-bibliography.bbl}

\end{document}